\documentclass[a4paper,10pt,twoside]{cpc-hepnp}

\usepackage{multicol}
\usepackage{graphicx}
\usepackage{booktabs}
\usepackage{amssymb,bm,mathrsfs,bbm,amscd}
\usepackage[tbtags]{amsmath}
\usepackage{lastpage}
\usepackage{CJK}

\begin{document}
\begin{CJK*}{GBK}{song}



\title{Probing the mass degeneracy of particles with different spins\thanks{Supported by National Natural Science
Foundation of China (Nos. 11347124, 11147003, U1204115) and Doctoral Scientific Research Foundation, University of South China.}}

\author{%
      ZHANG Zhen-Hua $^{1;1)}$\email{zhangzh@iopp.ccnu.edu.cn}%
\quad L\"{U} Gang $^{2;2)}$\email{ganglv@haut.edu.cn}%
\quad WEI Ke-Wei $^{3;3}$ \email{weikw@hotmail.com}
}
\maketitle

\address{%
$^1$ School of Nuclear Science and Technology, University of South China, Hengyang, Hunan 421001, China\\
$^2$ College of Science, Henan University of Technology, Zhengzhou 450001, China\\
$^3$ College of Physics and Electrical Engineering, Anyang Normal University, Anyang, Henan 455002, China
}

\begin{abstract}
The spin is an important property of a particle.
Although it is unlikely to happen, there is still a possibility that two particle with different spins share similar masses.
In this paper, we propose a method to probe this kind of mass degeneracy of particles with different spins.
We will use the cascade decay $B^+\to X(3872)K^+$, $X(3872)\to D^+D^-$ to explain our method.
It can be seen that the possible mass degeneracy of $X(3872)$ can lead interesting behavior in the corresponding cascade decay.
\end{abstract}




\begin{multicols}{2}

It is always important to decide the spin of a new particle once it is discovered.
However, the statistic of the events corresponding to the new particle is usually low.
As a result, sometimes we can only find that the spin of the new particle has several possibilities.
Consequently, it is possible that several particles with different spins share similar masses.
In this paper, we propose a method to confirm or exclude this possibility.

We will use the cascade decay, $B^\pm\to X(3872) K^\pm$, $X(3872) \to D^+D^-$,  as a example to explain our method. $X(3872)$ was first discovered by Belle Collaboration in the year of 2003 and was the first exotic hadron ever been discovered \cite{Choi:2003ue}.
At first, the analysis of the $X(3872)$ angular distributions in the decay to $J/\psi\pi^+\pi^-$ favors $\text{J}^{\text{PC}} = 1^{++}$, or $2^{-+}$ \cite{Abulencia:2006ma,Choi:2011fc}.
Only in last year, the LHCb Collaboration found that the spin of  $X(3872)$ is 1 \cite{Aaij:2013zoa}.
Before the discovery of the LHCb Collaboration, there is a possibility that  two particles with spin-1 and -2 degenerate around 3872 MeV.
We will show that this kind of mass degeneracy will lead some interesting behavior.

To probe the degeneracy of $X(3872)$,
we assume that there are two particles with spin-1 and -2, respectively, and with masses about 3872 MeV.
We will denote these two particles as $X_1$ and $X_2$, respectively.

When the invariant mass of $D^+D^-$ pair lies around 3872 MeV, the decay amplitude $\mathcal{M}$ for the cascade decay can be expressed as \cite{Zhang:2013iga}
\begin{equation}
\mathcal{M}(s_{12}, s_{13})= a_{1}P_1\big(g_{s_{D\overline{D}}}(s_{DK})\big)+a_{2}P_2\big(g_{s_{D\overline{D}}}(s_{DK})\big),
\end{equation}
where $P_l$ ($l=1,2$) is the $(l+1)$-th Legendre polynomial, $a_lP_l$ represents the decay amplitude with $X_l$ being the intermediate  resonance, $s_{D\overline{D}}$ and $s_{DK}$ are the invariant mass squared of $D^+D^-$ pair and $D^+$ and $K^+$ [$D(\overline{D})$ in the subscript represents $D^+(D^-)$], respectively.
The function $g_{s_{D\overline{D}}}(s_{DK})$ is defined as
\begin{equation}
g_{s_{D\overline{D}}}(s_{DK})=(\hat{s}_{DK}-s_{DK})/\Delta_{DK},
\end{equation}
where $\hat{s}_{DK}=(s_{DK}^{\text{max}}+s_{DK}^{\text{min}})/2$,  $\Delta_{DK}=(s_{DK}^{\text{max}}-s_{DK}^{\text{min}})/2$, with $s_{DK}^{\text{max(min)}}$ being the kinematically allowed maximum (minimum) value of $s_{DK}$.
 The subscript $s_{D\overline{D}}$ of the function $g_{s_{D\overline{D}}}(s_{DK})$ indicates its dependence on $s_{D\overline{D}}$ (through $s_{DK}^{\text{max}}$ and $s_{DK}^{\text{min}}$).
 The definition of $s_{DK}^{\text{max}}$ and $s_{DK}^{\text{min}}$ can be found in Ref. \cite{Zhang:2013iga}.

The Legendre Polynomials $P_{1}(x)$ and $P_2(x)$ have one and two zero points, which is (are) $x_0=0$ and $x_\pm=\pm1/\sqrt{3}$, respectively.
Correspondingly, the zero point(s) for  $P_{1}\big(g_{s_{DD}}(s_{DK})\big)$ and $P_{2}\big(g_{s_{DD}}(s_{DK})\big)$ are $s_{DK}^{(0)}=\hat{s}_{DK}$ and $s_{DK}^{(\pm)}=\hat{s}_{DK}\pm\Delta_{DK}/\sqrt{3}$, respectively.
This allows us to divide the allowed region of $s_{DK}$, $(s_{DK}^{\text{min}},s_{DK}^{\text{max}})$, into four regions, which are $(s_{DK}^{\text{min}},s_{DK}^{(-)})$, $(s_{DK}^{(-)},s_{DK}^{(0)})$, $(s_{DK}^{(0)},s_{DK}^{(+)})$, and $(s_{DK}^{(+)}, s_{DK}^{\text{max}})$, and will be denoted as $\Omega_a$, $\Omega_b$, $\bar{\Omega}_b$, and $\bar{\Omega}_a$, respectively.
As was shown in Refs. \cite{Zhang:2013iga} and \cite{Zhang:2013oqa}, localized $CP$ asymmetry can be used to probe the interference of two resonances with different spins.
However, for the cascade decay that we are considering, the weak phase is $\arg(V_{tb}V_{ts}^{\ast}/V_{cb}V_{cs}^\ast)$, which is too small to have any detective effects.
In stead, we will propose some other quantities to probe the degeneracy of $X(3872)$.

Since the weak phase is small, we can neglect it safely.
Then, we can redefine $a_l$ ($l=1,2$) according to
\begin{equation}
a_l\rightarrow a_le^{i\delta_l},
\end{equation}
so that the new defined $a_l$'s are real, and $\delta_l$'s are strong phases.
The absolute value squared of $\mathcal{M}$ can be expressed as
\begin{equation}
|\mathcal{M}|^2=\mathcal{S}+\mathcal{A}
\end{equation}
where
\begin{eqnarray}
\mathcal{S}\!&=&\!\frac{a_1^2}{\Delta_{DK}^2}\big(s_{DK}^{(0)}-s_{DK}\big)^2\nonumber\\
&&\!
+\frac{9a_2^2}{4\Delta_{DK}^4}\big(s_{DK}^{(-)}-s_{DK}\big)^2\big(s_{DK}^{(+)}-s_{DK}\big)^2,\\
\mathcal{A}\!&=&\! \frac{3a_1a_2\cos\delta}{\Delta_{DK}^3} \!\big(\!s_{DK}^{(0)}\!\!-\!s_{DK}\!\big)\!\big(\!s_{DK}^{(-)}\!\!-\!s_{DK}\!\big)\!\big(\!s_{DK}^{(+)}\!\!-\!s_{DK}\!\big),
\end{eqnarray}
with $\delta=\delta_2-\delta_1$.
There is a very interesting property for $\mathcal{S}$ and $\mathcal{A}$.
One can see from the above two equations that with the replacement:
\begin{equation}
s_{DK}\rightarrow 2s_{DK}^{(0)}-s_{DK},
\end{equation}
$\mathcal{A}$ changes its sign, while $\mathcal{S}$ does not:
\begin{eqnarray}
\mathcal{S}(s_{DK})&=&\mathcal{S}(2s_{DK}^{(0)}-s_{DK}),\\
\mathcal{A}(s_{DK})&=&-\mathcal{A}(2s_{DK}^{(0)}-s_{DK}).
\end{eqnarray}
In other words, as a function of $x=(s_{DK}-\hat{s}_{DK})/\Delta_{DK}$, $\mathcal{S}$ is even, while $\mathcal{A}$ is odd.
In Fig. \ref{Fig}, we depict $\mathcal{S}$, $\mathcal{A}$ and their sum as a function of $x$ for $a_1=a_2=\cos\delta=1$.
The four regions in phase space for fixed $s_{D\overline{D}}$ that were mentioned above now correspond to $x\in (-1,1/\sqrt{3})$, $(-1/\sqrt{3},0)$, $(0,1/\sqrt{3})$, and $(1/\sqrt{3},1)$, respectively.

Inspired by the above property, we define two quantities $R_a$ and $R_b$ as
\begin{equation}
R_r=\frac{N_{\Omega_r}-N_{\bar{\Omega}_r}}{N_{\Omega_r}+N_{\bar{\Omega}_r}},
\end{equation}
where $r=a$ or $b$, $N_\omega$ ($\omega=\Omega_a, \Omega_b,\bar{\Omega}_b, \bar{\Omega}_a$) represents the event number of the cascade decay in the region $\omega$.
It can be seen for the situation we are considering that
\begin{equation}
R_r=\frac{\mathcal{A}_{\Omega_r}}{\mathcal{S}_{\Omega_r}},
\end{equation}
where
\begin{eqnarray}
\mathcal{S}_{\Omega_r}=\int_{\Omega_r}\text{d}s_{DK} \mathcal{S},~~~~
\mathcal{A}_{\Omega_r}=\int_{\Omega_r}\text{d}s_{DK} \mathcal{A}.
\end{eqnarray}
After doing the above integral, one has
\begin{eqnarray}
\mathcal{S}_{\Omega_a}&=&\bigg[\bigg(1-\frac{1}{3\sqrt{3}}\bigg)\frac{a_1^2}{3}+\bigg(1-\frac{2}{3\sqrt{3}}\bigg)\frac{a_2^2}{5}\bigg]\Delta_{DK},\\
\mathcal{S}_{\Omega_b}&=&\bigg(\frac{a_1^2}{3}+\frac{2a_2^2}{5}\bigg)\frac{\Delta_{DK}}{3\sqrt{3}},\\
\mathcal{A}_{\Omega_a}&=&\frac{1}{3}\Delta_{DK}a_1a_2\cos\delta,\\
\mathcal{A}_{\Omega_b}&=&-\frac{1}{12}\Delta_{DK}a_1a_2\cos\delta.
\end{eqnarray}
With the above equations, one can further derive the relations between $R_{a}$ and $R_{b}$:
\begin{equation}
\frac{R_{b}}{R_{a}}=-\frac{1}{8}\left[3\sqrt{3}-2+\frac{3\sqrt{3}}{1+\frac{6a_2^2}{5a_1^2}}\right],
\end{equation}
and
\begin{equation}
-\frac{1}{4}(3\sqrt{3}-1)<\frac{R_{b}}{R_{a}}<-\frac{1}{4}\left(\frac{3\sqrt{3}}{2}-1\right).
\end{equation}
It can be seen that if there are two resonances lying around 3872 MeV, we should observe nonzero $R_a$ and $R_b$ with opposite signs.
If there is only one resonance lying around 3872 MeV, this means $a_1$ or $a_2$ equals to zero.
By setting $a_1$ or $a_2$ equal to zero, one can immediately see that $\mathcal{A}=0$.
As a result, both $R_a$ and $R_b$ equal to zero.
Thus, by measuring $R_a$ and $R_b$, one can confirm or exclude the degeneracy of $X(3872)$.

Note that all the above discussions are for fixed $\sqrt{s_{D\overline{D}}}$, which is around 3872 MeV.
Practically, $\sqrt{s_{D\overline{D}}}$ takes values which are in a small interval around 3872 MeV.
Then the above discussion should be modified by taking into account proper integral over $s_{D\overline{D}}$.

The upper limit for the branching ratio of the cascade decay is (at confident level of 90\%) \cite{PDG2012,Abe:2003zv}
\begin{equation}
\mathcal{B}(B^+\to X(3872)K^+)\times\mathcal{B}(X\to D^+D^-)<4.0\times10^{-5}.
\end{equation}
Thus the upper limit for the cascade decay event number $N_{\text{cas}}$ is $N_{B^+}\times 4.0\times10^{-5}$, where $N_{B^+}$ is the total event number for the $B^+$ signal.
One can also include the $CP$ conjugate channel to double the statistics.
The method that we proposed above fails when the relative strong phase $\delta$ is close to $\pi/2$.

%
\begin{center}
\includegraphics[width=0.48\textwidth]{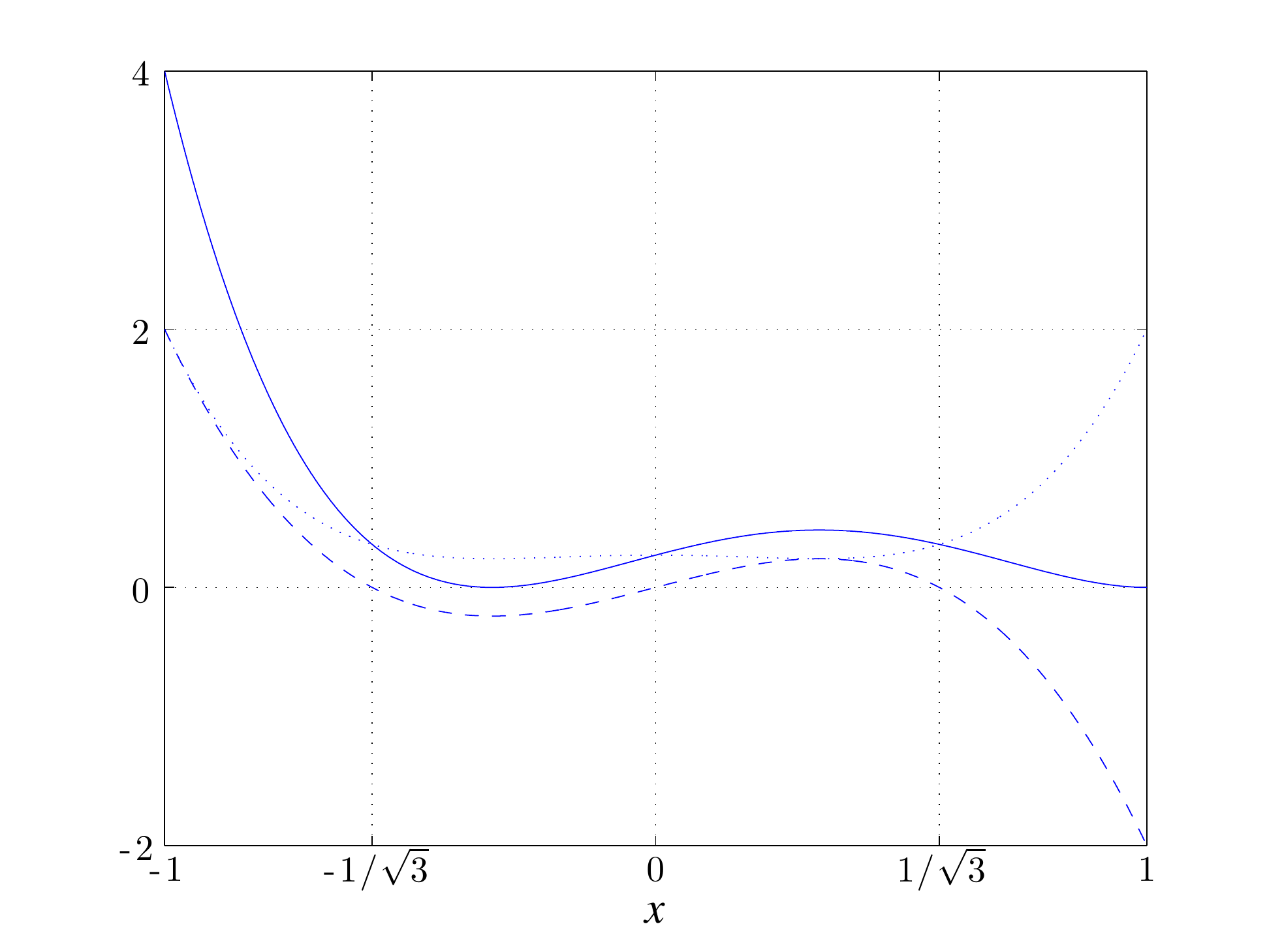}
\figcaption{\label{Fig}   $\mathcal{S}$ (dotted line), $\mathcal{A}$ (dashed line), and $|\mathcal{M}|^2$ (solid line) as a function of $x$, for the situation $a_1=a_2=\cos\delta=1$.
$(-1,-1/\sqrt{3})$, $(-1/\sqrt{3},0)$, $(0,1/\sqrt{3})$, and $(1/\sqrt{3}, 1)$ correspond to $\Omega_a$, $\Omega_b$, $\bar{\Omega}_b$, and $\bar{\Omega}_a$, respectively. }
\end{center}
%

%


\end{multicols}

\vspace{10mm}

\vspace{-1mm}
\centerline{\rule{80mm}{0.1pt}}
\vspace{2mm}

\begin{multicols}{2}

\end{multicols}

\clearpage

\end{CJK*}
\end{document}